\documentclass[twocolumn,showpacs,amsmath,amssymb]{revtex4}
\usepackage{bm}
\begin{document}
\draft

\newcommand{\pp}[1]{\phantom{#1}}
\newcommand{\be}{\begin{eqnarray}}
\newcommand{\ee}{\end{eqnarray}}
\newcommand{\ve}{\varepsilon}
\newcommand{\vs}{\varsigma}
\newcommand{\Tr}{{\,\rm Tr\,}}
\newcommand{\pol}{\frac{1}{2}}
\newcommand{\ba}{\begin{array}}
\newcommand{\ea}{\end{array}}
\newcommand{\bear}{\begin{eqnarray}}
\newcommand{\eear}{\end{eqnarray}}
\title{
Quantum Aspects of Semantic Analysis and Symbolic Artificial Intelligence
}
\author{Diederik Aerts $^1$ and Marek Czachor $^2$\\
$^1$ Centrum Leo Apostel (CLEA) and Foundations of the Exact Sciences (FUND)\\
Vrije Universiteit Brussel, 1050 Brussels, Belgium\\
$^2$ Katedra Fizyki Teoretycznej i Metod Matematycznych\\
Politechnika Gda\'nska, 80-952 Gda\'nsk, Poland}

\begin{abstract}
Modern approaches to semanic analysis if reformulated as Hilbert-space problems reveal formal structures known from quantum mechanics. Similar situation is found in distributed representations of cognitive structures developed for the purposes of neural networks. 
We take a closer look at similarites and differences between the above two fields and quantum information theory. 
\end{abstract}
\maketitle

\section{Prologue}

Let us consider an arbitrary text written by means of a 16-letter alphabet, say: a, b, c, $\dots$, n, o, p. Let us regroup as large part of the text as possible in quadruples belonging to the set $Q=\{$aeim, afim, agim, $\dots$, dhlm, dhln, dhlo, dhlp$\}$, and formed by strings obtained by picking out a single letter from a row of the matrix
\be
\left[
\begin{array}{cccc}
{\rm a} & {\rm b} & {\rm c} & {\rm d}\\
{\rm e} & {\rm f} & {\rm g} & {\rm h}\\
{\rm i} & {\rm j} & {\rm k} & {\rm l}\\
{\rm m} & {\rm n} & {\rm o} & {\rm p}
\end{array}
\right]
\ee
when one moves downwards starting from the first row. 
Now let us define the functions $F$ and $G$ by
$F({\rm a})=F({\rm d})=F({\rm e})=F({\rm h})=F({\rm i})=F({\rm l})=F({\rm m})=F({\rm p})=+1$, 
$F({\rm b})=F({\rm c})=F({\rm f})=F({\rm g})=F({\rm j})=F({\rm k})=F({\rm n})=F({\rm o})=-1$, $G(x_1x_2x_3x_4)=F(x_1)+F(x_2)+F(x_3)-F(x_4)$. On each four-character string of the regrouped part of the text we evaluate the value of $G$ and take its average value 
$\langle G\rangle$. 

The above awkward-looking manipulation with the text is an example of a procedure one might find in a paper on quantitative linguistics or semantic analysis. The analysis reveals certain correlational or contextual aspects of the text, the role of the contextuality measure being played by the average $\langle G\rangle$. 

To see what kind of a correlation one can capture, let us parametrize the alphabet by primed and unprimed bits 0, 1, $0'$, $1'$:

a~$=(00)$, b~$=(01)$, c~$=(10)$, d~$=(11)$, 

e~$=(00')$, f~$=(01')$, g~$=(10')$, h~$=(11')$, 

i~$=(0'0)$, j~$=(0'1)$, k~$=(1'0)$, l~$=(1'1)$, 

m~$=(0'0')$, n~$=(0'1')$, o~$=(1'0')$, p~$=(1'1')$.

After the reparametrization the regrouped text might represent data of an experiment testing the Bell inequality \cite{Bell} and the function $F$ represents values of the Bell observable for a single pair of measurements. And vice versa, any result of an experiment that tests the Bell inequality can be represented as a text written in a 16-letter alphabet.

The result of the form $|\langle G\rangle|> 2$ reveals a nonclassical probabilistic structure behind the text. This structure is, of course, typical of the 
{\it source\/} of the text, since the text itself may be a simple collection of characters on a computer printout. Actually, we can immediately identify the nonclassical elements disclosed by $|\langle G\rangle|> 2$: The bits $0$ and $0'$ (or $1$ and $1'$) correspond to {\it nonorthogonal\/} vectors, and ordered pairs such as $(01)$ are represented by tensor products. The possibility of hiding information behind nonorthogonal bases is the key idea of quantum cryptography \cite{BB84,E91} and tensor representations of conjuctions are fundamental to quantum information theory (QIT). The observation of Bell that correlations between symbols in ``texts" may reveal the presence of nonorthogonal bases is perhaps the most ingenious ingredient of his famous paper \cite{Bell}.

The idea that some sort of mathematical manipulation with texts, or some apparently artificial mathematical representation of them, may reveal deep structures such as similarity of meaning or other nontrivial correlations, is at the roots of semantic analysis (SA). Still another field where analogies with the Bell inequality example are particularly striking is related to neural-network distributed representations of concepts \cite{Plate}. The links of such scientific disciplines with quantum mechanics, and QIT in particular, are almost unexplored as yet. The present paper is an attempt of filling up the gap \cite{q-ph}.

\section{Vector models of texts}

Modern approaches to SA typically model words and their meanings by vectors from finite-dimensional vector spaces. The prominent examples of such approaches 
are  Latent Semantic Analysis (LSA) \cite{SDDFLH90,LFL98}, Hyperspace Analogue to Language 
(HAL) \cite{Lund}, Probabilistic Latent Semantic Analysis (pLSA) \cite{pLSA}, Latent Dirichlet Allocation \cite{LDA}, Topic Model \cite{TM}, or Word Association Space (WAS) \cite{Steyvers}.
In the present Letter we concentrate on a simplified version of LSA, but we believe the discussion we present can be applied to all vector models of language and concept representation.

SA is typically based on text co-occurence matrices and data-analysis technique employing singular value decomposition (SVD). Various models of SA provide powerful methods of determining similarity of meaning of words and passages by analysis of large text corpora. 
The procedures are fully automatic and allow to analyze texts by computers without an involvment of any human understanding. For example, what makes LSA quite impressive comes from the experiments with simulation of human performance. LSA-programmed machines were able to pass multiple-choice exams such as Test of English as a Foreign Language (TOEFL) 
(after training on general English)  \cite{LD97} or, after learning from an introductory psychology textbook, a final exam for psychology students \cite{LFL98}. 

These and other achievements of LSA raise the question of its relevance for the problem of brain functioning and AI \cite{L02}. However, an element we found particularly intriguing and which is the main topic of our paper, is in similarities between LSA and formal structures of QIT.

LSA is essentially a Hilbert space formalism. One represents words by vectors spanning a finite-dimensional space and text passages are represented by linear combinations of such words, with appropriate weights related to frequency of occurence of the words in the text. Similarity of meaning is represented by scalar products between certain word-vectors 
(beloging to the so-called semantic space).

In QIT, words, also treated as vectors, are being processed by quantum algorithms or encoded/decoded by means of quantum cryptographic protocols. Although one starts to think of quantum programming languages \cite{Oemer,Bettelli,Seilinger}, the semantic issues of quantum texts  are difficult to formulate. LSA is in this context a natural candidate as a starting point for ``quantum linguistics".

Still, LSA has certain conceptual problems of its own. As stressed by many authors, the greatest difficulty of LSA is that it treats a text passage as a ``bag of words", a set where order is irrelevant \cite{LLF98}. 
The difficulty is a serious one since it is intuitively clear that syntax is important for evaluation of text meaning. The sentences ``Mary hit John" and 
``John hit Mary" cannot be distinguished by LSA; ``Mary did hit John" and ``John did not hit Mary" have practically identical LSA representations because ``not" is in LSA a very short vector \cite{L02}. What LSA can capture is that the sentences are about violence. 

We think that experience from QIT may prove useful here. A basic object in QIT is not a word but a letter. Typically one works with the binary alphabet consisting of 0 and 1 and qubits. Ordering of qubits is obtained by means of the tensor product.
Ordering of words can be obtained in the same way, but before we proceed with QIT formalism, let us explain the standard LSA and formulate it in quantum mechanical notation. 

\section{Semantic analysis: An illustration}

Let us consider the following passage:

``($s_1$)
How much wood would a woodchuck chuck if a woodchuck could chuck wood?
($s_2$)
Woodchuck would chuck as much wood as a woodchuck could chuck if a woodchuck could chuck wood. 
($s_3$)
Could woodchuck chuck 35 cubic feet of dirt?
($s_4$)
If a woodchuck could chuck wood woodchuck would chuck 700 pounds of wood."

The LSA matrix representation of this text reads
$
\begin{array}{rcccc}
                &    s_1 &    s_2 &    s_3 &    s_4 \\
{\rm how}       &     1  &     0  &     0  &     0  \\
{\rm much}      &     1  &     1  &     0  &     0  \\
{\rm wood}      &     2  &     2  &     0  &     2  \\
{\rm would}     &     1  &     1  &     0  &     1  \\
{\rm a}         &     2  &     2  &     0  &     1  \\
{\rm woodchuck} &     2  &     3  &     1  &     2  \\
{\rm chuck}     &     2  &     3  &     1  &     2  \\
{\rm if}        &     1  &     1  &     0  &     1  \\
{\rm could}     &     1  &     2  &     1  &     1  \\
{\rm 35}        &     0  &     0  &     1  &     0  \\
{\rm cubic}     &     0  &     0  &     1  &     0  \\
{\rm feet}      &     0  &     0  &     1  &     0  \\
{\rm of}        &     0  &     0  &     1  &     1  \\
{\rm dirt}      &     0  &     0  &     1  &     0  \\
{\rm 700}       &     0  &     0  &     0  &     1  \\
{\rm pounds}    &     0  &     0  &     0  &     1  \\
\end{array}
\to
$
$
A_0=
\left(
\begin{array}{cccc}
1  &     0  &     0  &     0  \\
1  &     1  &     0  &     0  \\
2  &     2  &     0  &     2  \\
1  &     1  &     0  &     1  \\
2  &     2  &     0  &     1  \\
2  &     3  &     1  &     2  \\
2  &     3  &     1  &     2  \\
1  &     1  &     0  &     1  \\
1  &     2  &     1  &     1  \\
0  &     0  &     1  &     0  \\
0  &     0  &     1  &     0  \\
0  &     0  &     1  &     0  \\
0  &     0  &     1  &     1  \\
0  &     0  &     1  &     0  \\
0  &     0  &     0  &     1  \\
0  &     0  &     0  &     1  \\
\end{array}
\right).
$
\medskip
\noindent

It is usual to pre-process $A_0$ by multiplying each entry by a function associated with the entropy of an appropriate word evaluated on the basis of an entire text. The question of what kind of a co-occurence matrix should one relate to a text is actually an open one, and is investigated in various alterantives to LSA (HAL, WAS, Topic Model). For simplicity we skip this point.

The text corresponds now to the map $A: \bm R^4\to\bm R^{16}$, whose SVD (up to numerical roundup errors) is $A_0=U^{\dag} D_0 V$ where  
\be
U^{\dag}
&=&
\left(
\begin{array}{cccc} 
-0.06 & -0.12 &  0.15 &  0.70\\
-0.14 & -0.15 &  0.35 &  0.08\\
-0.40 & -0.22 & -0.26 &  0.23\\
-0.20 & -0.11 & -0.13 &  0.11\\
-0.34 & -0.26 &  0.21 &  0.20\\
-0.50 &  0.11 &  0.04 & -0.20\\
-0.50 &  0.11 &  0.04 & -0.20\\
-0.20 & -0.11 & -0.13 &  0.11\\
-0.30 &  0.23 &  0.17 & -0.32\\
-0.02 &  0.37 &  0.11 &  0.18\\
-0.02 &  0.37 &  0.11 &  0.18\\
-0.02 &  0.37 &  0.11 &  0.18\\
-0.07 &  0.41 & -0.36 &  0.20\\
-0.02 &  0.37 &  0.11 &  0.18\\
-0.05 &  0.04 & -0.48 &  0.02\\
-0.05 &  0.04 & -0.48 &  0.02
\end{array}
\right),\\
D_0
&=&
\left(
\begin{array}{cccc}
8.38 & 0 & 0 & 0\\
0 & 2.52 & 0 & 0\\
0 & 0 & 1.79 & 0\\
0 & 0 & 0 & 1.04
\end{array}
\right)
,\\
V
&=&
\left(
\begin{array}{cccc} 
-0.52 & -0.67 & -0.17 & -0.48\\
-0.30 & -0.07 &  0.94 &  0.10\\
 0.28 &  0.34 &  0.21 & -0.86\\
 0.73 & -0.64 &  0.18 &  0.02
\end{array}
\right).
\ee
The essential step of LSA is the reduction 
\be
A_0=U^{\dag} D_0 V\mapsto A_1=U^{\dag} D_1 V
\ee
where $D_1= PD_0$ and $P$ is a projector commuting with $D_0$. For example, if 
\be
P
&=&
\left(
\begin{array}{cccc} 
1 & 0 & 0 & 0\\
0 & 0 & 0 & 0\\
0 & 0 & 0 & 0\\
0 & 0 & 0 & 0
\end{array}
\right)
\ee
then 
\be
A_1
=
\left(
\begin{array}{cccc} 
0.26 & 0.33 & 0.08 & 0.24\\
0.61 & 0.78 & 0.19 & 0.56\\
1.74 & 2.24 & 0.56 & 1.60\\
0.87 & 1.12 & 0.28 & 0.80\\
1.48 & 1.90 & 0.48 & 1.36\\
2.17 & 2.80 & 0.71 & 2.01\\
2.17 & 2.80 & 0.71 & 2.01\\
0.87 & 1.12 & 0.28 & 0.80\\
1.30 & 1.68 & 0.42 & 1.20\\
0.08 & 0.11 & 0.02 & 0.08\\
0.08 & 0.11 & 0.02 & 0.08\\
0.08 & 0.11 & 0.02 & 0.08\\
0.30 & 0.39 & 0.09 & 0.28\\
0.08 & 0.11 & 0.02 & 0.08\\
0.21 & 0.28 & 0.07 & 0.20\\
0.21 & 0.28 & 0.07 & 0.20      
\end{array}
\right).
\ee
We will not go very deeply into details of how and why a reduced representation, of the type illustrated by $A_1$, may allow a computer to pass TOEFL not worse than an average non-native speaker who wants to study in the USA, and refer the reader to publications on LSA. For our purposes it is sufficient to know that
the rows of $A_1$ are termed the word-vectors and the space of word-vectors is known as the semantic space. Cosines between two word-vectors (or just their scalar products) are measuring a semantic distance (similarity of meaning) between words within a given set of text corpora represented by $A$. What is important, SVD can make some entries of $A_1$ negative and even make some scalar products negative, the latter occuring for antonyms. The coefficients of word-vectors lose, after SVD, the simple link to frequncies of occurences of words. 

Of course, the dimensions appearing in real texts investigated by means of LSA are much greater 
(for example 30473 columns and 60768 rows in the experiment discussed in \cite{LD97}). Experience shows that the analysis is most efficient if the projector $P$ projects on a subspace of dimension around 300, but what is the meaning of this dimension is yet a subject of speculations \cite{WAS}. 

\section{Semantic analysis in quantum notation}

In our example the matrix $U^{\dag}$ is not square but its columns are mutually orthogonal. 
Taking any 12 orthonormal vectors that are, in addition, orthogonal to the columns of $U^{\dag}$ we can replace $U^{\dag}$ by a $16\times 16$ unitary matrix $\tilde U^{\dag}$ whose first 4 columns coincide with those of $U^{\dag}$, and end up with SVD of the form
\be
\tilde A_k=\Big(A_k,0\Big)= 
\tilde U^{\dag}
\left(
\begin{array}{cc}
D_k & 0\\
0 & 0
\end{array}
\right)
\left(
\begin{array}{cc}
V & 0\\
0 & V^\perp
\end{array}
\right)
=\tilde U^{\dag}
\tilde D_k
\tilde V
,\nonumber
\ee
$k=0,1$, where all the matrices are square and $V^\perp$ is an arbitrary unitary matrix of appropriate dimension. The map $A_k\mapsto \tilde A_k$ neither adds nor removes any information from the text; its only objective is to work with text matrices and their SVDs that may be regarded as operators mapping certain Hilbert space $\cal H$  into itself. 

The Hilbert space ${\cal H}$ is finite dimensional, but in principle one cannot impose any limitation on the number of words or sentences one wants to take into account. It is therefore natural to treat all the concrete examples as subspaces of an infinite dimensional Hilbert space of all the possible words. Whether sentences or other text units are regarded as collections of words or as new words is a matter of convention. Assume each word of a vocabulary is represented by a basis vector $|n\rangle$, where $n$ is a natural number. The text matrix ($\tilde A=\tilde A_0$ or $\tilde A=\tilde A_1$) corresponds to the operator 
$
\hat A=\sum_{mn}A_{mn}|m\rangle\langle n|.
$
The column representing a $n$th sentence is given by the (unnormalized) vector 
\be
| s_n\rangle=\hat A |n\rangle=\sum_m A_{mn}|m\rangle.\label{psi_n}
\ee
For example, the sentence $s_2$ is in LSA represented by the sentence-vector
\be
| s_2\rangle 
&=&
|2\rangle
+
|4\rangle
+
|8\rangle
+
2\Big(
|3\rangle
+
|5\rangle
+
|9\rangle\Big)
+
3\Big(
|6\rangle
+
|7\rangle\Big).\nonumber
\ee
After SVD the coefficients of a sentence-vector are typically neither natural nor positive.
Let us note that $| s_2\rangle$ is not a word-vector in the sense of LSA, but a sentence-vector: Word-vectors are the {\it rows\/} of the text matrix. The rows are obtained from $\hat A$ by 
$
\langle  w_m|
=\langle m|\hat A.
$
The similarity of meaning of, say, ``how" and ``much" is given by
$
\cos({\rm how,much})
=
\langle  w_1| w_2\rangle/\big(\parallel  w_1\parallel\cdot \parallel  w_2\parallel\big).
$
(Recall that LSA gives optimal characterization of meaning if one calculates the scalar product after the reduction 
$D_0\mapsto D_1=PD_0$ with appropriately chosen $P$; in the example, before reduction 
$\cos({\rm how,much})
=0.707107$ and after the reduction $\cos({\rm how,much})=0.999985$). 

Putting this differently, the word-vectors characteristic of a text represented by the operator $\hat A$ are given by $| w_m\rangle=A^{\dag}|m\rangle$.
The matrix representing similarities of meaning between all the possible pairs of words corresponding to the text $\hat A$ is thus given by 
\be
\cos(m{\rm th\,word},n{\rm th\,word})
=
\frac{\langle m|\hat A\hat A^{\dag}|n\rangle}
{\sqrt{\langle m|\hat A\hat A^{\dag}|m\rangle}\sqrt{\langle n|\hat A\hat A^{\dag}|n\rangle}}.
\nonumber
\ee
As we can see, the entire information about mutual relations between words is in LSA encoded in the operator $\rho=\hat A\hat A^{\dag}$. Taking into account (\ref{psi_n}) and the resolution of unity $\bm 1=\sum_n |n\rangle\langle n|$ we can write
\be
\rho
=
\hat A\sum_n |n\rangle\langle n|\hat A^{\dag}
=
\sum_n| s_n\rangle\langle  s_n|
=
\sum_n p^s_n|\sigma_n\rangle\langle  \sigma_n|,
\label{rho}
\ee
with $p^s_n=\langle  s_n|s_n\rangle$ and $\langle  \sigma_n|\sigma_n\rangle=1$.
Since in any practical application the number of words is finite, the sum in (\ref{rho}) is finite as well and $\Tr \rho=\parallel A\parallel^2_{\rm HS}=
\sum_n\lambda_n<\infty$, where $\lambda_n$ are eigenvalues of $N=\hat A^{\dag}\hat A$, and 
$\parallel \cdot\parallel_{\rm HS}$ is the Hilbert-Schmidt norm. For this reason $\rho$ is formally an unnormalized density matrix of the set of sentences. 

The operator $N$ plays an essential role in LSA. To see this let us look at the explicit proof of SVD formulated in the quantum notation (physicists will recognize here the so-called Schmidt decomposition). Let $|\lambda_n\rangle$ be a normalized eigenvector of $N$, i.e. $N|\lambda_n\rangle=\lambda_n|\lambda_n\rangle$.  Denoting 
$|\alpha_n\rangle=\hat A |\lambda_n\rangle$ we 
compute
\be
\hat A 
&=&
\sum_{|\alpha_n\rangle\neq 0}|\alpha_n\rangle\langle\lambda_n|\nonumber\\
&=&
\sum_{|\alpha_n\rangle\neq 0}\frac{|\alpha_n\rangle}{\parallel \alpha_n\parallel}
\sqrt{\lambda_n}
\langle\lambda_n|\nonumber\\
&=&
\underbrace{
\sum_{k}|\beta_k\rangle\langle k|}_{\tilde U^{\dag}}
\underbrace{
\sum_{l}\sqrt{\lambda_l}|l\rangle\langle l|}_{\tilde D}
\underbrace{
\sum_{m}|m\rangle\langle\lambda_m|}_{\tilde V}
\ee
where $|\beta_k\rangle=|\alpha_k\rangle/\parallel \alpha_k\parallel$ if $\lambda_k>0$, or any other basis vector from the subspace corresponding to $\lambda_k=0$, if $\lambda_k=0$. It is clear that the singular values in SVD are given by $\sqrt{\lambda_k}$. The LSA procedure is essentially equivalent to the spectral analysis of  $N$. 

Let us finally note that $N$ can be written as 
\be
N=\hat A^{\dag}\sum_n|n\rangle\langle n|\hat A
=
\sum_n| w_n\rangle\langle  w_n|
=
\sum_n p^w_n|\omega_n\rangle\langle  \omega_n|,\nonumber
\ee
with $p^w_n=\langle  w_n|w_n\rangle$ and $\langle  \omega_n|\omega_n\rangle=1$,
i.e. as an unnormalized density matrix representing a mixture of word-vectors. 

\section{Supersymmetry and dimensional reductions}

The duality between sentence-vectors and word-vectors whose one of the manifestations 
is the link 
$
\hat A\hat A^{\dag}\leftrightarrow 
\hat A^{\dag}\hat A
$
is well known from supersymmetric theories \cite{SUSY}. In supersymmetric terminology 
operators $\hat A\hat A^{\dag}$ and $\hat A^{\dag}\hat A$
are known as superpartners. 

The dimensional reduction employed in LSA is performed on the spectrum of $N$. Since one eliminates in this way small eigenvalues, the procedure is analogous to some sort of purification of word-vector density matrices. But we know that one of the standard results of supersymmetric quantum mechanics states that $N$ and $\rho$  are isospectral. The interchange of $N$ and $\rho$ is equivalent to replacing word-vectors by sentence-vectors. Dimensional reduction can be thus performed for both $N$ and $\rho$, in the latter case the reduction deals with sentence-vector density matrices. Finally, one can combine the two approaches. 
A  ``supersymmetric LSA" can be based on supercharges  
$
Q=
\left(
\begin{array}{cc}
0 & A\\
A^{\dag} & 0
\end{array}
\right)
$
and the two density matrices taken simultaneously in $H=Q^2=\rho\oplus N$. 

In addition to the above dimensional reductions, two additional reductions are very natural from the viewpoint of our quantum interpretation. Let us note that in addition to the spectrum $\{\lambda_n\}$, we have two sets of ``mixing parameters": $\{p^s_n\}$ and 
$\{p^w_n\}$. The relations between them are the following 
\be
p^w_n
&=&
\langle  w_n|w_n\rangle
=\langle  n|AA^{\dag}|n\rangle
=\rho_{nn},\\
p^s_n
&=&
\langle  s_n|s_n\rangle
=\langle  n|A^{\dag}A|n\rangle
=N_{nn}.
\ee
Elimination of small diagonal elements $\rho_{nn}$ or $N_{nn}$ is not equivalent to eliminating small eigenvalues of $N$ or $\rho$. However, after this type of ``purification" 
the resulting operators $\tilde \rho$ and $\tilde N$ are still positive and, hence, can be factorized as $\tilde \rho=BB^{\dag}$, $\tilde N=C^{\dag}C$, leading effectively to two new types of reduction: $A\mapsto B$ and $A\mapsto C$. 

\section{Fock space of words}

As we have seen, LSA can be formulated as a Hilbert space problem. The ``bag of words" analysis is performed in $\cal H$. Ordered sequences of words can, in principle, be constructed in exact analogy to ordered sequences of letters in QIT. Still, there is a subtlety we want to point out. 

Consider a phrase, i.e. an ordered $n$-tuple of words, 
$({\rm word}_1,\dots, {\rm word}_n)$. Quantum physicist's intuition tells us that the natural representation of the sentence is a tensor product of vectors representing the words. The difficulty is this: Which vectors should one choose? The mutually orthogonal basis vectors 
$|j_1\rangle,\dots, |j_n\rangle$, or rather the associated word-vectors 
$|w_1\rangle=A^{\dag}|j_1\rangle,\dots, |w_n\rangle=A^{\dag}|j_n\rangle$?

Whatever representation one chooses, the phrase $(n_1,\dots,n_K)$ will be mapped into
\be
|n_1\dots n_K\rangle=|n_1\rangle\otimes\dots\otimes|n_K\rangle
\in \overbrace{{\cal H}\otimes\dots\otimes {\cal H}}^K={\cal H}^{\otimes K}\nonumber.
\ee
Including the empty word we arrive at the Fock space of all the text passages 
$
{\cal H}_F=\oplus_{K=0}^\infty {\cal H}^{\otimes K}.
$

LSA is performed in ${\cal H}_F$ in exactly the same way as in ${\cal H}$. 
The structures one can investigate are much richer. Taking as an example G.~Stein's  
phrase ``Rose is a rose is a rose is a rose", not only can we work with
\be
| s_1\rangle
=
4|{\rm rose}\rangle
+
3|{\rm is}\rangle
+
3|{\rm a}\rangle\in
\cal H
\ee
but also with vectors revealing the syntactic structures, for example, 
\be
| s_2\rangle
&=&
|{\rm rose}\rangle
\oplus
3|{\rm is}\rangle
\otimes
|{\rm a}\rangle
\otimes
|{\rm rose}\rangle
\in
{\cal H}\oplus {\cal H}^{\otimes 3}\subset {\cal H}_F,\nonumber\\
| s_3\rangle
&=&
\Big(|{\rm rose}\rangle
+
3|{\rm is}\rangle\Big)
\oplus
3
|{\rm a}\rangle
\otimes
|{\rm rose}\rangle
\in
{\cal H}\oplus {\cal H}^{\otimes 2}\subset {\cal H}_F. \nonumber
\ee
The above formulas show a typical feature of Fock spaces, namely superpositions of vectors belonging to different tensor powers. It is very interesting that similar constructions are encountered in convolution-based memory models, such as TODAM \cite{TODAM} or Holographic Reduced Representations (HRRs) 
\cite{Plate}.

\section{Relation to Smolensky's tensor product binding}

Smolensky in \cite{Smolensky} proposed tensor products of vectors as a means of solving the so-called binding problem: How to keep track of which features belong to which objects in a formal connectionist model of coding? In the linguistic context of SA the binding problem is equivalent to the problem of representing syntax. Links to quantum structures are particularly striking here, but there are also intriguing logical differences with what one would expect from a QIT perspective. 

First, one represents an {\it activity state\/} of a network by a vector, and this is very close to what a quantum physicist would do. In comments to his Definition~2.1  Smolensky stresses that the vectors are always written in the same and fixed basis. So formally we do not really need vectors, but $n$-tuples of numbers are enough. This is against the philosophy of QIT where states are indeed vectors and the same information may be encoded in non-parallel vectors. 

The fact that preferred basis is used becomes even more important in models such as TODAM or HRRs where the tensor product is replaced by its ``compressed form": convolution or circular convolution. Both operations are defined on $n$-tuples and not on vectors. Still, one can argue that in quantum measurement theory we do indeed deal with preferred pointer bases \cite{Bush} and the models such as HRRs may refer to this level of analysis. 

A predicate $\texttt{p(a,b)}$, such as $\texttt{eat(John,fish)}$, is represented by the vector $\bm r_1\otimes \bm a
+\bm r_2\otimes \bm b$ where the vectors $\bm r_k$ represent {\it roles\/} and 
$\bm a$, $\bm b$ are {\it fillers\/}. A predicate is, accordingly, given by an {\it entangled activity state\/}. A person trained on QIT would expect the vector to mean ``role $\bm r_1$ AND filler $\bm a$, OR role $\bm r_2$ AND filler $\bm b$". Of course, the intention of Smolensky was different: The sum is meant to represent the conjuction (AND) and not the alternative (OR). This feature is also characteristic of other neural-network models. 
Why is it so and is this type of representation crucial for symbolic AI?

The above similarities and differences show that further exploration of possible implications of connectionist models for QIT, and vice versa, may be worth of further studies. We will not pursue these matters further here. 

\section{Efficiency of tensor representations}

Tensor products are more ``economic" than Cartesian powers due to the identifications 
of the type 
$
\big(\alpha|\psi\rangle\big) \otimes |\phi\rangle 
=
|\psi\rangle \otimes \big(\alpha|\phi\rangle \big)
=
\alpha\big(|\psi\rangle \otimes |\phi\rangle \big)
$
that do not hold in Cartesian products. Thus the Fock space automatically performs a kind of dimensional reduction, which is the main idea of both LSA and distributed representations. 

If we are more interested in the issue of binding than in ordering of words then further compression of information is possible if one employs symmetric (bosonic) or antisymmetric 
(fermionic) Fock spaces. Symmetric tensor powers are closer to convolutions employed in HRRs but, unlike convolutions, are defined on vectors and not $n$-tuples of numbers.

Let us also note that in binary (or qubinary) representations all tensor powers can be decomposed into irreducible components, exactly in the same way it is performed in 2-spinor calculus \cite{PR}. It is known that any irreducible representation corresponds to symmetric spinors and any antisymmetric spinor is a scalar times the singlet (all antisymmetric two-index spinors are proportional to one another). So it is very natural indeed to employ representations based on symmetric operations as the main building blocks of, say, memory models (convolution used in HRRs is also commutative).

All these links are interesting from the point of view of the discussions  between Penrose and proponents of classical AI \cite{Penrose vs AI}. 
If brain is a quantum device, as suggested in \cite{Penrose} or, which is a weaker condition,  if the conceptual part of the mind entails a formal quantum structure \cite{GA2002,Kybernetes}, then the presence of tensor structures in SA or AI will not be accidental. 

The question of tensor representations of semantic aspects of texts in principle can be settled experimentally. Document retrieval experiments based on quantum logic were already performed \cite{Widdows} and the results are encourageing. 

Let us finally make the remark that some authors stress (cf. \cite{Hampton82}) that semantic categorizations cannot be modelled by a set logic. Experiments were reported where, for instance, people were willing to accept that chairs are a type of furniture and that carseats are a type of chair, but would then deny that carseats are a type of furniture (for a review cf. \cite{PetFish}). Trying to model the meanings of `furniture', `chair', `carseat' by means of set-theoretical constructions one arrives at contradiction with the inequality $P(A\wedge B\wedge C)\leq P(A\wedge C)$ (cf. also \cite{Aerts}). In QIT this type of contradiction is at the roots of the Bell inequality violation, whose proof is based on set-theoretic constructions while QIT employs tensor structures in Hilbert spaces. Similarly, tests of tensor structures via SA may play an analogous role in AI, quantitative linguistics, or experimental psychology, as the Bell inequality did for hidden-variables theories.

\acknowledgments

This research was supported by Grant G.0339.02 of the Flemish Fund for Scientific Research and the KBN Grant No. PBZ-Min-008/P03/03. We are indebted to T. Plate for his comments on tensor structures in distributed representations.


\begin{references}
\bibitem{WAS}WAS is a memory vector model where performance comparable to LSA is obtained with only 20-40 dimensions.
\end{references}
\end{document}